\documentclass[conference,a4paper]{IEEEtran}

\usepackage[utf8]{inputenc}
\usepackage[T1]{fontenc}
\usepackage{microtype}
\usepackage[nocompress]{cite}
\usepackage{xspace}
\usepackage{relsize} % for \smaller

\usepackage{amsmath}
\usepackage{amssymb}
\usepackage{stackrel}

\usepackage{balance}
\usepackage{paralist}
\usepackage[lowtilde]{url}
\usepackage{enumitem}

\renewcommand{\subsection}[1]{\smallskip\noindent{\bf{#1.}}}
\newcommand{\smallsec}[1]{\smallskip\noindent{\bf\textit{#1.}}}
\renewcommand{\subsubsection}[1]{\smallskip\noindent{\em\bfseries #1.}}

\usepackage{longtable}
\usepackage{booktabs}
\usepackage{multirow}
\usepackage{rotating} % for rotated cells

\usepackage{tabulary}

\usepackage{dcolumn}
\newcolumntype{d}{D{.}{.}{2.1}}
\newcolumntype{e}{D{.}{.}{4.0}}
\newcolumntype{f}{D{.}{.}{4.1}}
\newcolumntype{g}{D{.}{.}{5.1}}
\newcolumntype{h}{D{.}{.}{2.2}}
\newcolumntype{i}{D{.}{.}{3.2}}
\newcolumntype{j}{D{.}{.}{4.2}}
\newcolumntype{k}{D{.}{.}{5.2}}
\newcolumntype{m}{D{.}{.}{1.2}}

\usepackage{caption}
\usepackage[table]{xcolor}
\usepackage{graphicx}
\graphicspath{{figures/}}

\usepackage{tikz}
\usetikzlibrary{arrows}
\usetikzlibrary{calc}
\usetikzlibrary{chains}
\usetikzlibrary{fadings}
\usetikzlibrary{fit}
\usetikzlibrary{positioning}
\usetikzlibrary{shapes}
\newcommand\tikzmark[1]{\tikz[remember picture,overlay]\node[inner ysep=0pt] (#1) {};}

\usepackage[noend]{algorithmic}
\usepackage{algorithm}

\usepackage{fancyvrb}
\usepackage{listings}
\lstdefinestyle{C}{
    language=C,
    basicstyle=\ttfamily\smaller,
    numberblanklines=true,
    columns=fixed,
    aboveskip=2pt,
    belowskip=1pt,
    lineskip=0pt,
    numbers=left,
    numberstyle=\tiny,
    numberfirstline=true,
    firstnumber=1,
    xleftmargin=15pt,
    morekeywords={assert},
}

\usepackage[rounded]{syntax}
\let\oldsdlengths\sdlengths
\renewcommand\sdlengths{\oldsdlengths\setlength{\sdfinalskip}{0pt}}

\usepackage[%
  pdfpagemode=UseNone,% No sidebar in pdf viewer
  pdfpagelabels=false,plainpages=true,% Work if the document has no page numbers
  ]{hyperref}

\makeatletter
\renewcommand\fs@ruled{\def\@fs@cfont{\bfseries}\let\@fs@capt\floatc@ruled
  \def\@fs@pre{\kern4pt\hrule height.8pt depth0pt \kern2pt}%
  \def\@fs@post{\kern0pt\hrule\relax\vspace{-5mm}}%
  \def\@fs@mid{\kern2pt\hrule\kern2pt}%
  \let\@fs@iftopcapt\iftrue}
\makeatother

\newcommand\definetool[2]{\newcommand{#1}{{\smaller\sc #2}\xspace}}
\definetool{\blast}     {Blast}
\definetool{\cpachecker}{CPAchecker}
\definetool{\cbmc}      {Cbmc}
\definetool{\cil}       {Cil}
\definetool{\llvm}      {Llvm}
\definetool{\tvla}      {Tvla}
\definetool{\ocaml}     {OCaml}
\definetool{\tvp}       {Tvp}
\definetool{\camplp}    {CamlP4}
\definetool{\foci}      {Foci}
\definetool{\tcp}       {TCP}
\definetool{\escjava}   {ESC/Java}
\definetool{\slam}      {Slam}
\definetool{\jpf}       {JPF}
\definetool{\sycmc}     {SyCMC}
\definetool{\impact}    {Impact}
\definetool{\wolverine} {Wolverine}
\definetool{\ufo}       {UFO}
\definetool{\java}      {Java}
\definetool{\scratch}   {Scratch}
\definetool{\esbmc}     {Esbmc}
\definetool{\pkind}     {\textsc{PKind}}
\definetool{\mathsat}   {\textsc{MathSAT5}}

\newcommand{\kinduction}{$k$\nobreakdash-induction\xspace}
\newcommand{\kInduction}{$k$\nobreakdash-Induction\xspace}

\newcommand{\true}{\mathit{true}}
\newcommand{\false}{\mathit{false}}

\newcommand{\cpa}{\mathbb{D}}

\newcommand{\merge}{\mathsf{merge}}

\newcommand{\reached}{\mathit{reached}}

\newcommand{\inv}{\mathit{Inv}}

\renewcommand{\implies}{\Rightarrow}

\newcommand{\pr}{\pi}

\newcommand{\mytitle}{Combining k-Induction with\\ Continuously-Refined Invariants}
\newcommand{\myauthor}{Dirk Beyer, Matthias Dangl, and Philipp Wendler}

\hypersetup{
	pdfauthor={\myauthor},
	pdftitle={\mytitle},
}

\begin{document}
% set page counter to 0, so that the first page with content
% gets page number 1
% setting the page counter to 1 for first page triggers some
% LaTeX warnings with references
\setcounter{page}{0}

\thispagestyle{empty}
\begin{minipage}{17cm}
\begin{center}
~\\[3cm]
\Huge{\mytitle}
\\[2cm]
\large{\myauthor}
\\[1cm]
\normalsize
{University of Passau, Germany}\\[7cm]

\hspace{-5mm}
\includegraphics[scale=0.2]{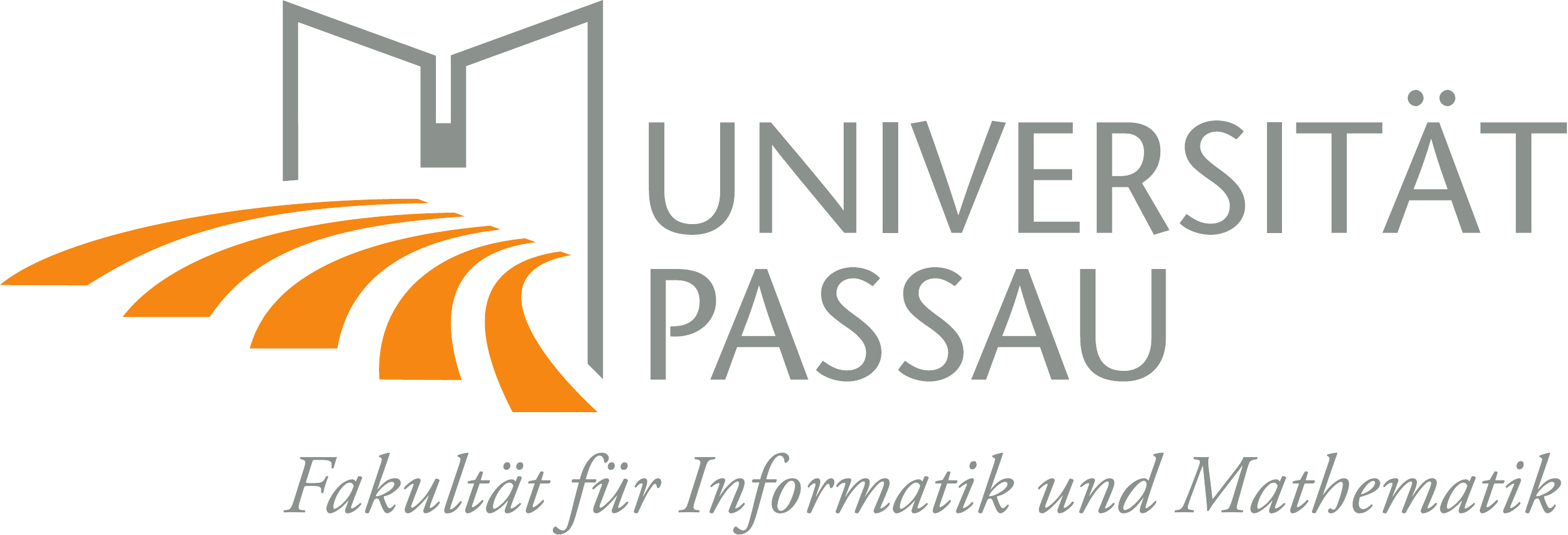} \\[1cm]
Technical Report, Number MIP-1503\\
Department of Computer Science and Mathematics\\
University of Passau, Germany\\
January 2015
\end{center}
\end{minipage}

\title{\mytitle}

\author{{\myauthor}
  \vspace{3mm}\\
  {University of Passau, Germany}\\
}
\maketitle
\thispagestyle{plain}
\pagestyle{plain}

\begin{abstract}
  Bounded model checking (BMC) is a well-known and successful technique for finding bugs in software.
\kinduction is an approach to extend BMC-based approaches from falsification to verification.
Automatically generated auxiliary invariants can be used to strengthen the induction hypothesis.
We improve this approach and further increase effectiveness and efficiency
in the following way:
we start with light-weight invariants
and refine these invariants continuously during the analysis.
We present and evaluate an implementation
of our approach
%of bounded model checking with \kinduction
in the open-source verification-framework \cpachecker.
Our experiments show that
combining \kinduction with continuously-refined invariants
significantly increases effectiveness and efficiency,
and outperforms all existing implementations of 
\kinduction-based software verification
in terms of successful verification results.

\end{abstract}

%\begin{keywords}
%Software Verification,
%Bounded Model Checking,
%\kInduction,
%Invariant Generation
%\end{keywords}

\section{Introduction}

%In the past, real-world verification was mostly limited to hardware verification.
Advances in software verification in the recent years
have lead to increased efforts towards applying formal verification methods
to industrial software, in particular operating-systems code~\cite{LDV,SLAMtransfer}.
%%%db: How does this paper support the claim?
%~\cite{Algorithms-for-Software-Model-Checking:-Predicate-Abstraction-vs.-Impact}.
One model-checking technique that is implemented by more than half of the verifiers
that participated in the 2014 Competition on Software Verification~\cite{SVCOMP14}
is bounded model checking (BMC)~\cite{Bounded-Model-Checking}.
For unbounded systems, BMC can be used only for falsification, 
not for verification~\cite{Handbook-of-Satisfiability}.
This limitation to falsification can be overcome
by combining BMC with mathematical induction
and thus extending it to verification~\cite{Automatic-Analysis-of-Scratch-pad-Memory-Code-for-Heterogenous-Multicore-Processors}.
Unfortunately, inductive approaches
are not always powerful enough
to prove the required verification conditions,
because not all program invariants are inductive~\cite{Automatic-Invariant-Strengthening-to-Prove-Properties-in-Bounded-Model-Checking}.
This problem can be mitigated by using the more general \kinduction
instead of the standard induction~\cite{Checking-Safety-Properties-Using-Induction-and-a-SAT-Solver},
an approach which has already been implemented in the DMA-race analysis tool \scratch~\cite{Automatic-Analysis-of-DMA-Races-Using-Model-Checking-and-k-Induction}
and in the software verifier \esbmc~\cite{ESBMC-COMP13}.
Nevertheless, additional supportive measures are often required
to guide \kinduction and take advantage of its full potential~\cite{Software-Verification-Using-k-Induction}.
Our goal is to provide a powerful and competitive approach
for reliable, general-purpose software verification
based on BMC and \kinduction,
implemented in a state-of-the-art software verification framework.

Our contribution is a new combination of
\kinduction-based model checking with
automatically-generated continuously-refined invariants
that are used to strengthen the induction hypothesis,
which increases the effectiveness of the approach.
BMC and \kinduction are combined
in an algorithm that iteratively increments the induction parameter~$k$.
The invariant generation runs in parallel to the \kinduction proof construction,
starting with relatively weak (but inexpensive to compute) invariants,
and increasing the strength of the invariants over time as long as the analysis continues.
The \kinduction-based proof construction adopts
the currently known set of invariants in every new proof attempt.
This approach can verify easy problems quickly
(with a small initial~$k$ and weak invariants),
and is able to verify complex problems by increasing the effort
(by incrementing~$k$ and searching for stronger invariants).
Thus, it is both efficient and effective.
In contrast to previous work~\cite{ESBMC-COMP13}, the new approach is sound.
We implemented our approach as part of the open-source 
software-verification framework \cpachecker~\cite{CPAchecker},
and we perform an extensive experimental comparison of our implementation against the two existing tools
that use similar techniques
and against another successful software-verification approach.

\subsection{Availability of Data and Tools}
Our experiments are based on benchmark verification tasks
from the 2015 Competition on Software Verification.
All benchmarks, tools, and results
of our evaluation are available on
a supplementary web page%
\,\footnote{\url{http://www.sosy-lab.org/~dbeyer/cpa-k-induction/}}.

\subsection{Contributions}
We make the following novel contributions:
We develop an approach
for providing continuously refined invariants to \kinduction
by using configurable program analysis with precision refinement.
We also present an extensive evaluation
where we compare various different approaches and implementations
against the implementation of our proposed approach
and show that our technique outperforms
other approaches to software verification with \kinduction.

%\mynote{db: It would be a good service for the reviewers to briefly mention the most important contributions explicitly.
%Maybe move this to after the expample.}

\begin{figure*}
%\figurerule
\begin{minipage}[t]{0.48\linewidth}
\begin{lstlisting}[style=C,firstline=6]
int main() {
  unsigned int x1 = 0, x2 = 0; int s = 1;

  while (nondet()) {
    if (s == 1) x1++;
    else if (s == 2) x2++;

    s++;
    if (s == 5) s = 1;

    if ((s == 1) && (x1 != x2)) {
      // Valid safety property
ERROR: return 1;
    }
  }
}
\end{lstlisting}
%\vspace{-2mm}
\caption{Safe example program \texttt{example-safe}, which cannot be proven with existing \kinduction-based approaches}
\label{fig:example-safe}
\end{minipage}
\hfill
\begin{minipage}[t]{0.48\linewidth}
\begin{lstlisting}[style=C,firstline=6]
int main() {
  unsigned int x1 = 0, x2 = 0; int s = 1;

  while (nondet()) {
    if (s == 1) x1++;
    else if (s == 2) x2++;

    s++;
    if (s == 5) s = 1;
  }

  if (s >= 4) {
    // Invalid safety property (s may be 4)
ERROR: return 1;
  }
}
\end{lstlisting}
%\vspace{-2mm}
\caption{Unsafe example program \texttt{example-unsafe}, where \esbmc produces a wrong proof}
\label{fig:example-unsafe}
\end{minipage}
\vspace*{-2mm}
\end{figure*}

\subsection{Example}
\label{example}
We illustrate the open problem of \kinduction that we address,
and the strength of our approach, on two example programs.
Both programs encode an automaton,
which is typical, e.g., for software that implements
a communication protocol.
The automaton has a finite set of states,
which is encoded by variable~\texttt{s},
and two data variables~\texttt{x1} and~\texttt{x2}.
There are some state-dependent calculations (lines~5 and~6 in both programs)
that alternatingly increment \texttt{x1}~and~\texttt{x2},
and a calculation of the next state (lines~8 and~9 in both programs).
The state variable cycles through the range from~1~to~4.
These calculations are done in a loop
with a non-deterministic number of iterations.
Both programs also contain a safety property
(the label \texttt{ERROR} should not be reachable).
The program \texttt{example-safe} in Fig.~\ref{fig:example-safe}
checks that in every fourth state,
the values of \texttt{x1}~and~\texttt{x2} are equal;
it satisfies the property.
The program \texttt{example-unsafe} in Fig.~\ref{fig:example-unsafe}
checks that when the loop exits,
the value of state variable~\texttt{s} is not greater or equal to $4$;
it violates the property.

First, note that the program \texttt{example-safe}
is difficult or impossible to prove with other software-verification approaches:
(1)~BMC cannot prove safety for this program
because the loop may run arbitrarily long.
(2)~Explicit-state model checking fails
because of the huge state space
(\texttt{x1} and \texttt{x2} can get arbitrarily large).
(3)~Predicate analysis with counterexample-guided abstraction refinement (CEGAR)
and interpolation is able to prove safety,
but only if the predicate $\mathit{x1} = \mathit{x2}$ gets discovered.
If the interpolants contain instead only predicates such as
$\mathit{x1} = 1$, $\mathit{x2} = 1$, $\mathit{x1} = 2$, etc., the analysis will not terminate.
Which predicates get discovered is hard to control
and usually depends on internal interpolation heuristics of the 
satisfiability-modulo-theory (SMT) solver.
(4)~Traditional $1$\nobreakdash-induction is also not able to prove the program safe
because the assertion is checked only in every fourth loop iteration
(when \texttt{s} is $1$).
Thus, the induction hypothesis is too weak
(the program state \texttt{s = 4}, \texttt{x1 = 0}, \texttt{x2 = 1}
is a counterexample for the step case in the induction proof).

Intuitively, this program should be provable
by \kinduction with a~$k$ of at least $4$.
However, for every~$k$, there is a counterexample
to the inductive-step case that refutes the proof.
For such a counterexample, set \texttt{s = }$-k$, \texttt{x1 = 0}, \texttt{x2 = 1}
at the beginning of the loop.
Starting in this state, the program would increment~\texttt{s} $k$~times (induction hypothesis)
and then reach \texttt{s = 1} with property-violating values of \texttt{x1}~and~\texttt{x2}
in iteration~$k+1$ (inductive step).
It is clear that~\texttt{s} can never be negative,
but this fact is not present in the induction hypothesis,
and thus the proof fails.
This illustrates the general problem of \kinduction-based verification:
safety properties often do not hold in unreachable parts of the state space of a program,
and \kinduction alone does not distinguish between reachable and unreachable parts of the state space.
If \esbmc with \kinduction analyzes program \texttt{example-safe},
the analysis ---as expected--- iteratively increments~$k$ and loops infinitely,
failing to prove safety.

This program could of course be verified more easily
if it were rewritten to contain a stronger safety property such as
$s \geq 1 \wedge s \leq 4 \wedge (s = 2 \Rightarrow \mathit{x1} = \mathit{x2}+1) \wedge (s \neq 2 \Rightarrow \mathit{x1} = \mathit{x2})$
(which is a loop invariant and allows a proof by 1-induction without auxiliary invariants).
However, our goal is to automatically verify real programs,
and programmers usually 
neither write down trivial properties such as $s \geq 1$
nor too complex properties such as $s \neq 2 \Rightarrow \mathit{x1} = \mathit{x2}$.

With our approach of combining \kinduction with invariants,
the program is proved safe with $k = 4$
and the invariant $s \geq 1$.
This invariant is easy to find automatically using an inexpensive static analysis,
such as an interval analysis.
For bigger programs, a more complex invariant might be necessary,
which might get generated at some point
by our continuous strengthening of the invariant.
Furthermore, stronger invariants can reduce
the $k$~that is necessary to prove a program.
For example, the invariant
$s \geq 1 \wedge s \leq 4 \wedge (s \neq 2 \Rightarrow \mathit{x1} = \mathit{x2})$
(which is still weaker than the full loop invariant above)
allows to prove the program with $k = 2$.
Thus, our strengthening of invariants can also shorten
the inductive proof procedure and lead to better performance.

\esbmc~\cite{ESBMC-COMP13}
tries to solve this problem of a too-weak induction hypothesis
by initializing only the variables of the loop-termination condition
to a non-deterministic value in the step case,
and initializing all other variables to their initial value in the program.
However, this approach is not strong enough for the program \texttt{example-safe}
and even produces a wrong proof (unsound result) for the program \texttt{example-unsafe}.
This second example program contains a different safety property about~\texttt{s},
which is violated.
Because the variable~\texttt{s} does not appear in the loop-termination condition,
it is not set to an arbitrary value in the step case as it should be,
and the inductive proof wrongly concludes that the program is safe
because the induction hypothesis is too strong.
\esbmc misses the bug in this program and claims it is correct.
Our approach does not suffer from this unsoundness,
because we only add invariants to the induction hypothesis
that the invariant generation had proven to hold.

\subsection{Related Work}
The use of auxiliary invariants is a common technique in software verification~\cite{Automatic-Generation-of-Invariants-and-Intermediate-Assertions},\cite{Incremental-Invariant-Generation-Using-Logic-Based-Automatic-Abstract-Transformers},\cite{InvGen-An-Efficient-Invariant-Generator},
and techniques combining abstract interpretation and SMT solvers also exist~\cite{UFO-COMP13}.
In most cases, the purpose is to speed up the analysis.
For \kinduction, however,
the use of invariants is crucial in making the analysis terminate at all
(cf. Fig.~\ref{fig:example-safe}).
There are several approaches to software verification
using BMC in combination with \kinduction.

\smallsec{Split-Case Induction}
We use the \textit{split-case \kinduction{}} technique~\cite{Automatic-Analysis-of-DMA-Races-Using-Model-Checking-and-k-Induction, Automatic-Analysis-of-Scratch-pad-Memory-Code-for-Heterogenous-Multicore-Processors},
where the base case and the step case are checked in separate steps.
Due to the fact that this technique is only able to handle one loop at a time,
another similarity to the approach of
the earlier versions of \scratch~\cite{Automatic-Analysis-of-DMA-Races-Using-Model-Checking-and-k-Induction}
is the transformation of programs with multiple loops
into programs with only one single monolithic loop
using a standard approach~\cite{Compilers:-Principles-Techniques-and-Tools}.
The alternative of recursively applying the technique to nested loops
is discarded by the authors of \scratch~\cite{Automatic-Analysis-of-DMA-Races-Using-Model-Checking-and-k-Induction},
because the experiments suggested it was less efficient
than checking the single loop that is obtained by the transformation.
\scratch also supports \textit{combined-case \kinduction{}}~\cite{Software-Verification-Using-k-Induction},
for which all loops are cut
by replacing them with $k$ copies each for the base and the step case,
and setting all loop-modified variables to non-deterministic values
before the step case.
That way, both cases can be checked at once in the transformed program
and no special handling for multiple loops is required.
%
%%% PW: Is the following actually relevant?
When using combined-case \kinduction,
\scratch requires loops to be manually annotated with the required $k$ values,
whereas its implementation of split-case \kinduction
supports iterative deepening of $k$
as in our implementation.
Contrary to \scratch,
we do not focus on one specific problem domain~\cite{Automatic-Analysis-of-DMA-Races-Using-Model-Checking-and-k-Induction, Automatic-Analysis-of-Scratch-pad-Memory-Code-for-Heterogenous-Multicore-Processors},
but want to provide a solution for solving a wide range
of heterogeneous verification tasks.

\smallsec{Auxiliary Invariants}
While both the split-case and the combined-case \kinduction supposedly succeed
with weaker auxiliary invariants
than for example the inductive invariant approach~\cite{Weakest-Precondition-of-Unstructured-Programs},
the approaches still do require auxiliary invariants in practice,
and the tool \scratch
requires these invariants to be annotated manually~\cite{Automatic-Analysis-of-DMA-Races-Using-Model-Checking-and-k-Induction, Software-Verification-Using-k-Induction}.
There are techniques for automatically generating invariants
that may be used to help inductive approaches to succeed~\cite{InvGenX,Property-Directed-Incremental-Invariant-Generation, Automatic-Invariant-Strengthening-to-Prove-Properties-in-Bounded-Model-Checking}.
These techniques, however,
are not guaranteed to justify their additional effort
by providing the required invariants on time,
especially if strong auxiliary invariants are required.
Based on previous ideas of supporting \kinduction
with invariants generated by lightweight static analysis~\cite{Strengthening-Induction-Based-Race-Checking-with-Lightweight-Static-Analysis},
we therefore strive to leverage the power of the \kinduction approach
to succeed with auxiliary invariants
generated by a static analysis
based on intervals.
However, to handle cases where it is necessary to invest more effort
into invariant generation,
we increase the precision of these invariants over time.
%
%%% Forward condition and bounding assertions are the same
%%% explicit-value analysis can be described in journal version
%Another application of \kinduction
%in combination with bounded model checking
%that is similar to ours is described in \cite{ESBMC-COMP13},
%but instead of their forward condition
%we use an explicit-value analysis and bounding assertions
%to check if the bounded model check covers all reachable states.
%
A verification tool using a strategy similar to ours is \pkind~\cite{PKind:-A-parallel-k-induction-based-model-checker, Incremental-Invariant-Generation-Using-Logic-Based-Automatic-Abstract-Transformers},
a model checker for Lustre programs based on \kinduction.
In \pkind, there is a parallel computation of auxiliary invariants,
where potential invariants derived by templates
are iteratively checked via \kinduction
and, if successful, added to the set of known invariants.
While this allows for strengthening the induction hypothesis over time,
the template-based approach lacks the flexibility that is
available to an invariant generator using dynamic precision refinement~\cite{CPAplus},
and the required additional induction proofs are potentially expensive.

\smallsec{Unsound Strengthening of Induction Hypothesis}
\esbmc does not require additional invariants for \kinduction,
because it assigns non-deterministic values only to the loop-termination condition variables
before the inductive-step case~\cite{ESBMC-COMP13}
and thus retains more information than our as well as the \scratch implementation~\cite{Automatic-Analysis-of-DMA-Races-Using-Model-Checking-and-k-Induction, Software-Verification-Using-k-Induction},
but \kinduction in \esbmc is therefore potentially unsound.
Our goal is to
perform a real proof of safety by removing all pre-loop information in the step case,
thus treating the unrolled iterations in the step case truly as "any $k$ consecutive iterations",
as is required for the mathematical induction.
Our approach counters this lack of information
by employing incrementally improving invariant generation.

\smallsec{Parallel Induction}
In \pkind, base case and step case are checked in parallel,
and the latest version of \esbmc, version 1.23,
supports parallel execution of
the base case,
the forward condition,
and the inductive-step case.
In contrast, our base case and inductive-step case
are checked sequentially,
while our invariant generation runs in parallel
to the aforementioned base- and step-case checks.

\vspace{\baselineskip}
\section{Background}

We briefly explain existing concepts that our approach uses.

\subsection{Programs}
We use the same notion of programs to describe the theoretical aspects of our ideas
as in previous work~\cite{CPA}.
The presentation of our work is restricted
to a simple imperative programming language
that contains only assume operations and assignments.
All variables are assumed to be integers%
\,\footnote{Our implementation is based on \cpachecker, which supports C programs.}.
\textit{Programs} are represented by control-flow automata.
A \textit{control-flow automaton} (CFA) consists of a set~$L$ of program locations,
modeling the program counter~$l$,
the initial program location~$l_0$, modeling the program entry,
and a set $G \subseteq L \times Ops \times L$ of control-flow edges,
each of which models the operation that is executed during the flow of control from one program location to another.
The variables that occur in operations from $Ops$ are contained in the set~$X$ of program variables.
In our presentation, we assume that each program contains at most one loop.
In our implementation,
we handle programs with multiple loops
by transforming all loops into a single monolithic loop~\cite{Compilers:-Principles-Techniques-and-Tools}.

\subsection{Configurable Program Analysis}
We use the concepts of \textit{configurable program analysis}~(CPA)\cite{CPA}
with dynamic precision adjustment~\cite{CPAplus}.
%(Due to space limitations, we restrict the presentation to the essential parts,
%and refer to reader to the literature for more details.)
A CPA defines an abstract domain and a transfer relation,
together with a merge operator to specify what happens at meet points in the control-flow
and a stop operator to specify the fixed-point conditions.
The software-verification framework \cpachecker allows plugging in CPAs as components,
and CPAs can be reused and combined,
such that common tasks like tracking the program counter or the call stack
do not need to be considered in every single analysis.
The CPA algorithm optionally merges (as defined by the merge operator)
newly-discovered abstract states 
with previously existing abstract states
to produce an abstract state covering both states,
over-approximating them.
This over-approximation may result in a loss of information,
but reduces the amount of states in favor of efficiency.
Each abstract state is paired with a precision, which specifies
how fine-grained the analysis should work
(to find a compromise between being efficient and precise).

\subsection{Bounded Model Checking}
The technique of \textit{bounded model checking} (BMC)~\cite{BMC} was
originally introduced as alternative to binary decision diagrams (BDD) 
in symbolic model checking,
%%%db: BDDs are also used for boolean decision procedures
% by using boolean decision procedures
to produce counterexamples more quickly,
and to speed up verification in general.
Classic BMC reduces model checking
to propositional satisfiability~(SAT):
Only counterexamples up to a given length~$k$ are considered
and a propositional formula~$f$ is constructed
such that $f$ is satisfiable iff such a counterexample exists.
A SAT~solver can be used to check the satisfiability of~$f$
and, if $f$ is satisfiable,
the counterexample can be reconstructed from the model for~$f$,
which is provided by the SAT~solver.
However, if $f$ is unsatisfiable,
no counterexample with a length smaller than or equal to~$k$ exists.
Thus, unless it is known that all reachable states are covered
by BMC with length~$k$,
the absence of longer counterexamples cannot be guaranteed.
Therefore, BMC is often classified as a technique for falsification,
not for verification.
Nowadays, BMC is based on solvers for satisfiability modulo theories~(SMT)~\cite{SMT-Based-Bounded-Model-Checking-for-Embedded-ANSI-C-Software}.

\subsection{k-Induction}
BMC-based approaches can be extended from falsification to verification
by induction.
Consider a program that contains a loop, and a safety property~$P$.
BMC with $k = 1$ may show that no counterexample
(a violation of $P$) of length $k = 1$ exists (a),
but a longer counterexample might still exist.
If, however, we are able to prove that
for any given iteration through the loop where $P$ holds before,
$P$~also holds after the iteration (b),
the program is verified by induction,
where (a) is the base case and (b) is the inductive-step case.
Consider as a more formal example
the standard induction principle over natural numbers:
\[
  \left(P(0) \land \forall n: \left( P(n) \implies P(n+1) \right) \right) \implies \forall n: P(n)
\]
This can be extended to greater values of~$k$
by asserting the safety property~$P$
for not only $1$ but $k$ consecutive predecessors in the step case,
which is known as \textit{\kinduction{}}.
\kinduction over natural numbers can be written as:
\[
  \!\left(\bigwedge^{k-1}_{i=0}\! P(i) \land \forall n: \left(\!\!\left(\bigwedge^{k-1}_{i=0}\! P(n\!+\!i) \!\right) \implies P(n\!+\!k)\!\!\right)\!\!\right)\! \implies \forall n: P(n)
\]
Intuitively, the induction proof is more likely to succeed
for higher values of~$k$,
because the inductive-step case asserts the safety property for more consecutive predecessors,
thus a less general case is checked.
It holds that for $k>1$, ${(k-1)}$-induction implies \kinduction
and that therefore $(k-1)$-induction must always be at least as hard as 
\kinduction~\cite{The-k-Induction-Principle}.

\subsection{Invariants}
An assertion~$p$ is called an \textit{invariant} of a program
if $p$ is true for all states of that program~\cite{Temporal-Verification-of-Reactive-Systems}.
If $p$ is an assertion that specifies the safety property of a program
and $p$ is invariant,
then the program is safe.
Proving the invariance of an assertion is therefore a method of software verification.
An assertion $\varphi$ is called \textit{inductive},
if it is provable by induction~\cite{Property-Directed-Incremental-Invariant-Generation}.
However, not every \textit{invariant} assertion is \textit{inductive}.
One solution to this problem
is trying to find an \textit{inductive} assertion
$\varphi$ that is stronger than~$p$,
i.e., $\varphi \implies p$.
Trivially, if $\varphi$ is invariant then $p$ is also invariant.
This strengthening of assertions can be achieved
by creating the conjunction of $p$
and an \textit{auxiliary invariant}~$p'$
such that $\varphi := p \land p'$\cite{Automatic-Invariant-Strengthening-to-Prove-Properties-in-Bounded-Model-Checking}.
%
%A very intuitive explanation of why this approach makes sense
%is given in~\cite{Automatic-Invariant-Strengthening-to-Prove-Properties-in-Bounded-Model-Checking}:
By choosing the auxiliary invariant in a way
that excludes those unreachable "good" states
that have transitions to "bad" successor states,
the stronger \textit{invariant} may be \textit{inductive}
where the weaker one was not.

\addtolength\textfloatsep{\baselineskip}
\vspace{\baselineskip}
\section{k-Induction with Invariants}

Our verification approach consists of two algorithms that run concurrently.
One algorithm is responsible for the generation of program invariants,
starting with imprecise invariants that are continuously refined (strengthened).
The other algorithm is responsible
for finding counterexamples with BMC
and constructing safety proofs with \kinduction,
for which it periodically picks up the invariants
that the former algorithm has constructed so far.
The \kinduction algorithm uses information from the invariant analysis,
but not vice versa.

\subsection{Iterative-Deepening k-Induction}
Algorithm \ref{k-induction-algo} shows our extension
of the \kinduction algorithm
to a combination with continuously-refined invariants.
Starting with an initial value for the bound $k$, e.g.,~$1$,
we iteratively increase the value of~$k$ after each unsuccessful attempt at
finding a specification violation or proving correctness of the program using \kinduction.
The following description of our approach to \kinduction
is based on split-case \kinduction~\cite{Software-Verification-Using-k-Induction},
where for the propositional state variables $s$ and $s'$
within a state transition system representing the program,
the predicate $I(s)$ denotes that $s$ is an initial state,
$T(s,s')$ states that a transition from $s$ to $s'$ exists,
and $P(s)$ asserts the safety property for the state $s$.

\smallsec{Base Case}
Lines~\ref{k-induction-algo-base-start} to~\ref{k-induction-algo-base-end} show the \textit{base case},
which consists of running BMC with the current bound~$k$.
This means that starting from an initial program state,
all paths of the program up to a maximum loop bound~$k$
are explored.
(As an optimization, one can omit checking for property violations
which have been checked in previous iterations with lower values of~$k$ already.)
Formally,
there exists a counterexample of length at most~$k$
if the following holds:
%
%\vspace{-1mm}
\[
  I(s_0) \land \bigvee_{n=0}^{k-1} \left(\bigwedge_{i=0}^{n-1} T(s_i, s_{i+1}) \land \lnot P(s_n)\right)
\]
If a counterexample is found, the algorithm terminates.

\smallsec{Forward Condition}
Otherwise we check
whether there exists a path with length $k' > k$ in the program,
or whether we have already fully explored the state space of the program (lines~\ref{k-induction-algo-fc-start} to~\ref{k-induction-algo-fc-end}).
In the latter case the program is safe and the algorithm terminates.
This check is called the \textit{forward condition}\cite{Proving-Transaction-and-System-level-Properties}.
Formally, the program was fully explored and is safe
if the following is unsatisfiable:
%
%\vspace{-1mm}
\[
  I(s_0) \land \bigwedge_{i=0}^{k-1} T(s_i, s_{i+1})
\]

\begin{algorithm}[t]
\caption{Iterative-Deepening \kInduction}
\label{k-induction-algo}
\begin{algorithmic}[1]
    \REQUIRE ~\\
             the initial value $k_{init} \geq 1$ for the bound $k$,\\
             an upper limit $k_{max}$ for the bound $k$,\\
             a function $\mathsf{inc}:\mathbb{N}\rightarrow\mathbb{N}$ with $\forall n\in\mathbb{N}: \mathsf{inc}(n) > n$\\\quad for increasing the bound $k$,\\
             the initial states defined by the predicate $I$,\\
             the transfer relation defined by the predicate $T$, and\\
             a safety property $P$\\
%             the function $\mathsf{sat}$ for checking the satisfiability of a formula
%             the function $\mathsf{get_currently_known_invariant}$ for polling the latest generated invariants

    \ENSURE \TRUE{} if $P$ holds, \FALSE{} otherwise

    \STATE $k := k_{init}$

    \WHILE{$k \leq k_{max}$}
    \STATE $\mathit{base\_case} := I(s_0) \land \bigvee\limits_{n=0}^{k-1} \left(\bigwedge\limits_{i=0}^{n-1} T(s_i, s_{i+1}) \land \lnot P(s_n)\right)$
    \label{k-induction-algo-base-start}
%      \STATE \hskip\algorithmicindent $I(s_1) \land T(s_1, s_2) \land \ldots \land T(s_{k-1}, s_k)$
%      \STATE \hskip\algorithmicindent $\land\ (\lnot P(s_1) \lor \ldots \lor  \lnot P(s_k))$
      \IF{$\mathsf{sat}(\mathit{base\_case})$}
         \label{invariant-gen-algo-safety-start}
        \RETURN \FALSE
      \ENDIF
      \label{invariant-gen-algo-safety-end}
      \label{k-induction-algo-base-end}
\vspace*{0.3\baselineskip}
        \STATE $\mathit{forward\_condition} := I(s_0) \land \bigwedge\limits_{i=0}^{k-1} T(s_i, s_{i+1})$
        \label{k-induction-algo-fc-start}
        \IF{$\lnot\,\mathsf{sat}(\mathit{forward\_condition})$}
          \RETURN \TRUE
        \ENDIF
        \label{k-induction-algo-fc-end}
\vspace*{0.4\baselineskip}
        \STATE $\begin{aligned}\mathit{step\_case}_n :=& \bigwedge\limits_{i=n}^{n+k-1} \left(P(s_i) \land T(s_i, s_{i+1})\right)\\ &  \land \lnot P(s_{n+k})\end{aligned}$
        \label{k-induction-algo-ind-start}
        \REPEAT
          \STATE $\inv := \mathsf{get\_currently\_known\_invariant}()$ \tikzmark{inv-consumer}
\vspace*{0.2\baselineskip}
          \IF{$\lnot\,\mathsf{sat}(\exists n \in \mathbb{N}: \inv(s_n) \land \mathit{step\_case}_n)$}
            \label{k-induction-algo-ind-sat}
            \RETURN \TRUE
\vspace*{0.2\baselineskip}
          \ENDIF
        \UNTIL{$\inv = \mathsf{get\_currently\_known\_invariant}()$}
        \label{k-induction-algo-repeat-ind}
        \label{k-induction-algo-ind-end}
\vspace*{0.5\baselineskip}
       \STATE $k$ := $\mathsf{inc}(k)$
    \ENDWHILE
    \RETURN \textbf{unknown}
\end{algorithmic}
\vspace{1mm}
\end{algorithm}

\smallsec{Inductive Step}
Checking the forward condition can, however,
only prove safety for programs with finite (and short) loops.
Therefore the algorithm also attempts an inductive proof (lines~\ref{k-induction-algo-ind-start} to~\ref{k-induction-algo-ind-end}).
The base case for induction was already checked before.
The \textit{inductive-step case} checks that,
after any sequence of $k$~loop iterations without a property violation,
there is also no property violation in loop iteration~$k+1$.
For model checking of software, however, this would often fail.
The reason for this is that by induction
we try to prove the property for every part of the state space of the program.
Typically, a program has large parts of the state space that are unreachable,
for which the property might not hold
but which are irrelevant for the safety of the program.
As an example, a typical loop in a program
uses a loop counter which has only positive values,
and with induction we would try to prove the property
for all possible values of the loop counter,
including negative values.
The key to success for using induction for safety proofs of programs
is thus to exclude as many unreachable parts of the state space
as possible from the proof.
This can be done by adding assumptions about program variables
to the induction hypothesis.
In our approach, we make use of the fact
that the invariants that were generated so far
by the concurrently-running invariant-generation algorithm hold,
and conjunct these facts to the induction hypothesis.
%It is sufficient to assume the invariants only for the program states
%in the beginning of the sequence,
%because our invariants are inductive by construction.
%$\inv(s_n) \land T(s_n,s_{n+1})$ implies $\inv(s_{n+1})$ for all~$n$.
Thus, the inductive-step case can prove a program as safe if the following is unsatisfiable:
%
%\vspace{-1mm}
\[
  \exists n \in \mathbb{N}: \inv(s_n) \land \bigwedge\limits_{i=n}^{n+k-1} \left(P(s_i) \land T(s_i, s_{i+1})\right)  \land \lnot P(s_{n+k})
\]
where $\inv$ are the currently available program invariants.
If this formula is satisfiable,
the induction check is inconclusive,
and the program cannot be proved as safe or unsafe
with the current value of~$k$ and the current invariants.
If during the time of the satisfiability check of the step case
a new (stronger) invariant has become available
(condition in line~\ref{k-induction-algo-repeat-ind} is false),
we immediately recheck the step case with the new invariant.
This can be done efficiently using an incremental SMT solver
for the repeated satisfiability checks in line~\ref{k-induction-algo-ind-sat}.
Otherwise we start over with an increased value of~$k$.

Note that the inductive-step case is similar to BMC
that checks for the presence of counterexamples of exactly length~${k+1}$.
However, as the step case needs to consider any consecutive $k+1$ loop iterations,
and not only the first such iterations,
it does not assume that the execution of the loop iterations
begins in the initial state.
Instead, it assumes that there is a sequence of $k$ iterations
without any property violation
(this is the induction hypothesis).

\begin{algorithm}[t]
\caption{Continuous Invariant Generation}
\label{invariant-gen-algo}
\begin{algorithmic}[1]
    \REQUIRE ~\\
             a configurable program analysis with dynamic precision adjustment $\cpa$,\\
             the initial states defined by the predicate $I$,\\
	     a coarse initial precision $\pr_0$,\\
	     a safety property $P$\\

    \ENSURE \TRUE{} if $P$ holds

    \STATE $\pr := \pr_0$
    \STATE $\inv := \true$
    \LOOP
	\STATE $\reached := \mathsf{CPAAlgorithm}(\cpa, I(s), \pr)$
	\IF{$\forall s \in \reached : P(s)$}
	  \RETURN \TRUE
	\ENDIF
\vspace*{0.5\baselineskip}
	\STATE \tikzmark{inv-producer}$\inv := \inv \land \bigvee\limits_{s\,\in\,\reached} s$
	\label{invariant-gen-algo-inv}
\vspace*{0.5\baselineskip}
        \STATE $\pr := \mathsf{RefinePrec(\pr)}$
        \label{invariant-precision-refinement}
    \ENDLOOP
\end{algorithmic}
\end{algorithm}

\begin{tikzpicture}[remember picture,overlay]
	\draw[->,double,out=200,in=0] ($(inv-producer) + (-0.3em,0.3em)$) to ($(inv-consumer) + (-0.1em,0.3em)$);
\end{tikzpicture}

\subsection{Continuous Invariant Generation}
Our continuous invariant generation
incrementally produces stronger and stronger program invariants.
It is based on an invariant-generation procedure
that is run in a loop, each time with an increased precision.
Each time the invariant has been strengthened,
it can be used as auxiliary invariant by the \kinduction procedure.
It may happen that this analysis proves safety of the program all by itself,
but this is not its main purpose in our application.

\smallsec{Algorithm}
Algorithm \ref{invariant-gen-algo} shows our continuous invariant generation.
%\mynote{TODO: based on existing work}
The initial program invariant is represented by the formula $true$.
We start with running the invariant-generation analysis once with a coarse initial precision.
After each run of the program-invariant generation,
we strengthen the previously-known program invariants
with the newly-generated invariants (line~\ref{invariant-gen-algo-inv})
and announce it globally (such that the \kinduction algorithm can use it).
If the analysis was able to prove safety of the program,
the algorithm terminates
(lines~\ref{invariant-gen-algo-safety-start} to~\ref{invariant-gen-algo-safety-end}).
Otherwise, the analysis is restarted with a higher precision.
%Currently, this increase of precision is not counterexample-guided
%(i.e., not \cegar~\cite{ClarkeCEGAR}),
%although we wish to integrate counterexample guidance for the refinement in the future.

Our approach works with any kind of invariant generation procedure,
as long as its precision, i.e., its level of abstraction, is configurable.
We use the reachability algorithm $\mathsf{CPAAlgorithm}$
for configurable program analysis with dynamic precision adjustment~\cite{CPAplus}.
It takes as input a configurable program analysis~(CPA),
an initial abstract state, and a precision.
It returns a set of reachable abstract states
that form an over-approximation of the reachable program state.
This algorithm works with any abstract domain
that can be formalized as a CPA.
Depending on the used CPA and the precision,
the analysis done by $\mathsf{CPAAlgorithm}$
can be efficient and abstract like data-flow analysis
as well as expensive and precise like model checking.

\smallsec{Abstract Domain}
For the invariant generation
we use an abstract domain based on expressions over intervals.
Note that this is not a requirement of our approach,
which works with any kind of domain.
Our choice is based on the high flexibility of this domain,
which can be fast and efficient as well as precise.

The analysis is formalized and implemented as a~CPA~\cite{CPA}
with dynamic precision adjustment~\cite{CPAplus}.
An abstract state of our invariant-generation domain consists of
a mapping $M : X \rightarrow Expr$ from program variables to arithmetic expressions,
where $Expr$ is the set of expressions
and $X$ is the set of variables.
The set $Expr$ of expressions 
consists of binary expressions, unary expressions, program variables,
and disjunctions of intervals,
and is defined recursively as
$Expr \subseteq ((Expr \times B \times Expr) \cup (U \times Expr) \cup X \cup I)$,
where 
$B$ is the set of supported binary operators
$B = \{+, *, /, \%, =, <, $\textasciicircum$, |, \lor, \&, \land, \gg, \ll, \cup\}$,
$U$ is the set of supported unary operators
$U = \{\lnot, \sim, -\}$, and
$I$ is the set of 
disjunctions of intervals of the form~$[u, l]$ with $u, l \in \mathbb{Z} \cup \infty$.
The disjunctions of intervals
allow for an efficient representation of ranges,
and, unlike in single-interval-based approaches, gaps between ranges can also be represented.

\smallsec{Precision}
In our CPA, the precision is a triple~$(Y, n, w)$,
where $Y \subseteq X$ is a specific selection of important program variables,
$n$~is the maximal nesting depth of expressions in the abstract state,
and $w$~is a boolean specifying whether widening should be used.
Those variables that are considered important
will not be over-approximated by merging abstract states.
With a higher nesting depth,
more precise relations between variables can be represented.
The use of widening ensures timely termination
(at the expense of a lower precision)
even for programs with loops with many iterations,
like those in the examples~\ref{fig:example-safe} and \ref{fig:example-unsafe}.

\smallsec{Merge}
Our CPA merges two abstract states %$e_1$ and~$e_2$
if both states do not differ in the expressions that are stored
for the important program variables from the set~$Y$ of the precision.
%i.e., if $\forall v\in Y:e_1(v) = e_2(v)$.
This way, the loss of information resulting from merging two abstract states does not affect
the selected variables in~$Y$.
Naturally, the more variables are in the precision,
the fewer merges occur, resulting in a more precise but slower analysis.
To guarantee timely termination of the analysis
even over loops with many iterations,
like those shown in the examples \ref{fig:example-safe} and \ref{fig:example-unsafe},
a widening strategy for over-approximating
variable values may be used when merging abstract states.
Formally, for two abstract states~$e_1, e_2$ and a precision~$\pr = (Y,n,w)$
the merge operator is defined as
\[\merge(e_1, e_2, \pr) = \begin{cases}
	\mathsf{widen}(e_1, e_2) & \text{if } w \land \neg\mathsf{differ}_\pr(e_1,e_2)\\
	\mathsf{union}(e_1, e_2) & \text{if } \neg w \land \neg\mathsf{differ}_\pr(e_1,e_2)\\
	e_2 & \text{if } \mathsf{differ}_\pr(e_1,e_2)
\end{cases}\]
with $\mathsf{differ}_\pr(e_1,e_2) = \exists v\in Y:e_1(v) \not= e_2(v)$.
The operation $\mathsf{union}(e_1, e_2)$ returns an abstract state
where for each variable the union of the values for this variable in~$e_1$ and~$e_2$ is used.
The operation $\mathsf{widen}(e_1, e_2)$ over-approximates
by assigning to each variable only a single (potentially infinite) interval.

\smallsec{Precision Refinement}
The initial precision~$(\emptyset, 1, \true)$ for this analysis
specifies an empty set of variables as important variables,
i.e., abstract states belonging to the same program location
are always merged (by applying widening).
The maximum expression-nesting depth of~$n=1$
means that abstract states map program variables to a single variable or
to a disjunction of intervals (no arithmetic operators allowed).

Our main refinement strategy is to add variables to the set~$Y$ of important program variables,
first adding one variable,
and then doubling the size of the set in each refinement step.
When choosing variables for this step,
we visit the control-flow automaton backwards from the error location and pick variables
that appear in assume edges,
such that variables appearing in conditions close to the error location
get added first.
This refinement strategy is property-guided,
rather than counterexample-guided like CEGAR.

Additionally, we have a refinement step
that increments the expression-nesting depth to~$2$,
allowing more complex expressions,
such as an addition of a variable with a disjunction of intervals;
this refinement is helpful if an invariant
$x = y + 1$ is required, but the values of~$x$ and~$y$ cannot
be over-approximated precisely enough.
The third refinement strategy is to disable the use of widening.
Thus, the precision and the efficiency of the analysis is dynamically adjusted during 
the analysis.
The maximal precision we use for our CPA is~$(X,2,\false)$
which tracks all program variables almost fully precisely.
Of course, any other precision-refinement strategy applicable for the chosen CPA
can be used for our continuous invariant generation, too.
\addtolength\textfloatsep{-\baselineskip}

\pagebreak
\newcommand\nPrograms{2\,814\xspace}
\section{Experimental Evaluation}
We compare our approach
with other \kinduction-based approaches
implemented in the same tool
as well as with other \kinduction-based tools.

\subsection{Benchmark Verification Tasks}
%%%db: not the programs (= verifiers) but the verification tasks
As benchmark set we use verification tasks from
the 2015 Competition on Software Verification (SV-COMP'15)~\footnote{\url{http://sv-comp.sosy-lab.org/2015/}}.
We took all \nPrograms verification tasks from the categories
\textit{ControlFlow},
\textit{DeviceDrivers64},
\textit{HeapManipulation},
\textit{Sequentialized},
and \textit{Simple}.
The remaining categories were excluded
because they use features
(such as bitvectors, concurrency, and recursion)
that not all configurations of our evaluation support.
742~verification tasks in the benchmark set contain a known specification violation.
Although we cannot expect an improvement for these verification tasks when using auxiliary invariants,
we did not exclude them because this would unfairly benefit our approach
(which spends some effort generating invariants
which are not helpful when proving existence of a counterexample).

\subsection{Experimental Setup}
All experiments were conducted on computers with
two 2.6\,GHz 8-Core~CPUs (Intel Xeon E5-2560~v2) with 135\,GB of RAM.
The operating system was Ubuntu~14.04 (64~bit),
using Linux~3.13 and OpenJDK~1.7.
Each verification task was limited to
two CPU cores,
a CPU run time of 15\,min
and a memory consumption of 15\,GB.

\subsection{Presentation}
All benchmarks, tools, and the full results
of our evaluation are available on
a supplementary web page%
\,\footnote{\url{http://www.sosy-lab.org/~dbeyer/cpa-k-induction/}}.

All reported times are rounded to two significant digits.
We use the scoring scheme of SV-COMP'15
to calculate a score for each configuration.
For every real bug found, 1~point is assigned,
for every correct safety proof, 2 points are assigned.
A score of 6 points is subtracted for every wrong alarm (false positive) reported by the tool,
and 12 points are subtracted for every wrong proof of safety (false negative).
This scoring scheme values proving safety higher than finding counterexamples,
and significantly punishes wrong answers,
which is in line with the community consensus~\cite{SVCOMP14}
on difficulty of verification vs. falsification
and importance of correct results.
We consider this a good fit for evaluating an approach
such as \kinduction,
which targets at producing safety proofs.

In Fig.~\ref{fig:experiments-quantile-cputime-scores-tools}~and~\ref{fig:experiments-quantile-cputime-scores-other},
we present experimental results using a plot
of quantile functions for accumulated scores
as introduced by the Competition on Software Verification~\cite{SVCOMP13},
which shows the score and CPU~time for successful results
and the score for wrong answers.
A data point $(x, y)$ of a graph means
that for the respective configuration
the sum of the scores of all wrong answers
and the scores for all correct answers with a run time of less than or equal to $y$~seconds
is~$x$.
For the left-most point $(x,y)$ of each graph, the $x$-value
shows the sum of all negative scores for the respective configuration
and the $y$-value shows the time for the fastest successful result.
For the right-most point $(x,y)$ of each graph, the $x$-value
shows the total score for this configuration, and the $y$-value shows the
maximal run time.
A configuration can be considered better,
the further to the right (the closer to $0$) its graph begins (fewer wrong answers),
the further to the right it ends (more correct answers),
and the lower its graph is (less run time).

\subsection{Comparison of \kinduction-based approaches}
To allow a meaningful evaluation of our approach,
we implemented it together with other existing approaches in the same tool.
We used
the \java-based open-source software-verification framework \cpachecker~\cite{CPAchecker},
which is available online%
\,\footnote{\url{http://cpachecker.sosy-lab.org}}
under the Apache~2.0 license.
For benchmarking, we used revision~15\,499 from the trunk of the \cpachecker repository,
with \mathsat{}\,\footnote{\url{http://mathsat.fbk.eu}} as SMT solver.
The \kinduction algorithm of \cpachecker
was configured to increment~$k$ by~$1$ after each try
(in Alg.~\ref{k-induction-algo},
$\mathsf{inc}(k) = k+1$).
The precision refinement of the continuous invariant generation
was configured to increment the number of important program variables
in the first, third, fifth, and any further precision refinements.
The second precision refinement increments the expression-nesting depth,
and the fourth precision refinement disables the widening operator.

We evaluated the following \kinduction-based configurations:
(1)~without any auxiliary invariants,
(2)~with statically-generated invariants of different precisions,
(3)~with unsound invariants using a reimplementation of the heuristic of \esbmc~\cite{ESBMC-COMP13},
(4)~with our new continuously-refined invariants.

The \kinduction-based configuration using no auxiliary invariants
is an instance of Alg.~\ref{k-induction-algo} where
$\mathsf{get\_currently\_known\_invariant}()$ always returns an empty set of invariants
and Alg.~\ref{invariant-gen-algo} does not run at all.

The configurations using statically-generated invariants
are also instances of Alg.~\ref{k-induction-algo}.
Here, Alg.~\ref{invariant-gen-algo} runs in parallel,
however, it terminates after one loop iteration.
We denote these configurations with triples~$(s, n, w)$
which represent the precision~$(Y, n, w)$ of the invariant generation
with $s$~being the size of the set of important program variables ($s = |Y|$).
The first of these configuration is~$(0,1,\true)$,
which means that no variables are in the set~$Y$ of important program variables
(i.e., all variables get over-approximated by the merge operator),
the maximum nesting depth of expressions in the abstract state is~$1$,
and the widening operator is used.
The second configuration is~$(16,2,\true)$,
which means that $16$~variables are in the set~$Y$,
the nesting depth of expressions in the abstract state is limited to~$2$,
and the widening operator is used.
The third configuration is~$(16,2,\false)$,
where $16$~variables are in the set~$Y$,
the maximum nesting depth of expressions in the abstract state is~$2$,
and the widening operator is not used.
These configurations were selected
because they represent some of the extremes
of the precisions used during dynamic invariant generation.
It is, however, impossible to cover every possible valid configuration
within the scope of this paper.

The heuristic of \esbmc is to preserve information
about variable values before the loop
to help the step-case check to succeed.
A sound technique for using pre-loop information in the step-case
is to havoc the loop-modified variables,
i.e., to remove all information about these loop-modified variables,
but keep all other information~\cite{Software-Verification-Using-k-Induction},
effectively propagating constants to the step case.
\esbmc, however, heuristically selects only those variables for havocing
that appear in loop-termination conditions~\cite{ESBMC-COMP13}.
This technique is easier and computationally cheaper than generating sound auxiliary invariants,
but may lead to wrong verification results,
as shown in Sec.~\ref{example} for our Example~\ref{fig:example-unsafe}.

\begin{table}
\centering
\caption{Results of \kinduction-based configurations in \cpachecker for~all~\nPrograms~verification tasks
with different approaches for generation of auxiliary invariants}
\label{tab:induction-results}
\begin{tabular}{@{}l@{\quad}ff@{~}f@{~}fff@{}}
\hline
Invariant & \multicolumn{1}{@{}c@{}}{\multirow{2}{*}{none}}      & \multicolumn{3}{c}{static}    & \multicolumn{1}{@{}c@{}}{\esbmc} & \multicolumn{1}{@{}c@{}}{cont.-} \\
generation    &     & \multicolumn{1}{@{}c@{}}{$(0,1,t)$} & \multicolumn{1}{@{}c@{~\,}}{$(16,2,t)$} & \multicolumn{1}{@{}c@{}}{$(16,2,f)$} & \multicolumn{1}{@{}c@{}}{heuristic} &\multicolumn{1}{@{}c@{}}{refined} \\
\hline
	Score \phantom{\Large X}&1\,557 &    3\,184 &    3\,263 &    3\,177 &        204 & \textbf{3\,464} \\
	Correct results         &1\,036 &    1\,852 &    1\,893 &    1\,849 &     1\,827 & \textbf{1\,981} \\
	Wrong proofs            &     2 &\textbf{1} &\textbf{1} &         2 &        263 &      \textbf{1} \\
	Wrong alarms            &    12 &        10 &        11 &         9 &          8 &      \textbf{7} \\
	CPU time (h)            &   400 &       200 &       190 &       200 &        140 &             170 \\
	Wall time (h)           &   380 &       150 &       130 &       120 &        130 &             100 \\
\hline
\multicolumn{6}{@{}l}{Times for correct results only: \phantom{\Large X}}\\
	CPU time (h)    &    7.1 &         14 &     15 &     13 & 8.8 &  17 \\
	Wall time (h)   &    5.7 &        8.4 &    8.9 &    7.6 & 6.9 & 9.4 \\
\hline
\multicolumn{6}{@{}l}{$k$-Values for correct safe results only: \phantom{\Large X}} \\
	Max. final $k$  &    101 &        101 &    101 &    119 & 120 &  88  \\
	Avg. final $k$  &    2.4 &        2.0 &    2.3 &    2.3 & 2.0 & 2.4  \\
\hline
\end{tabular}
%\vspace{-1mm}
\end{table}

%\begin{figure}
%\centering
%\includegraphics[width=\linewidth]{experiments/quantile-cputime-score-approaches.pdf}
%\vspace{-7mm}
%\caption{Quantile functions of $k$-induction-based configurations in \cpachecker
%for accumulated scores showing the CPU time for the successful results}
%\label{fig:experiments-quantile-cputime-scores-approaches}
%\vspace{-4mm}
%\end{figure}

\subsubsection{Score}
Using the unsound heuristic of \esbmc for invariant generation produces $263$~wrong proofs,
which shows that this is not a suitable approach for proving program safety.
In contrast, the few wrong proofs produced by the other configurations
are not due to conceptual problems,
but only due to incompletenesses in the analyzer's handling
of certain constructs such as unbounded arrays and pointer aliasing.

The configuration with no invariant generation
receives the second-lowest score of~$1\,557$,
and (as expected) can verify only $1\,036$~programs successfully,
producing more than $800$~results less than any of the configurations
that use sound auxiliary invariants.
This shows that it is indeed important in practice
to enhance \kinduction-based software verification with invariants.

The configurations using static invariant generation
produce~$1\,852$,~$1\,893$, and~$1\,849$ correct results
and achieve scores of~$3\,184$,~$3\,263$, and~$3\,177$ points, respectively.
These results are close to each other,
but improve upon the results of the plain \kinduction
without auxiliary invariants by a score of~$1\,600$ to~$1\,700$.
%Even though these configurations solve a similar number of programs,
%a closer inspection reveals that each of the configurations
%is able to correctly solve significant amounts of programs
%where both other configurations run into timeouts.

This observation explains the high score of~$3\,464$ points
achieved by our approach using continuously-refined invariant generation.
By combining the advantages of fast and coarse precisions
with those of slow but fine precisions,
it correctly solves~$1\,981$ verification tasks,
which is~$88$ more correct results
than the best of the chosen static configurations.
It is thus clearly the best of all evaluated
\kinduction-based approaches.

\subsubsection{Performance}
Table~\ref{tab:induction-results} shows that
the fastest configuration in terms of CPU time
is the unsound approach,
which is easily explained by the fact
that it often produces incorrect proofs
after analyzing a low number of loop iterations of the program.
Due to the vast amount of wrong results,
the speed of the approach can hardly be considered a success.

By far the highest amount of time is spent
by the configuration using no auxiliary invariants,
because for those programs that cannot be proved without auxiliary invariants,
the \kinduction procedure loops incrementing~$k$ until the time limit is reached.
For the sound configurations,
the wall times for the correct results correlate with the amount of correct results,
i.e., on average about the same amount of time is spent on correct verifications,
whether or not invariant generation is used.
This shows that the overhead of generating auxiliary invariants is well-compensated.

The configurations with static and continuously-refined invariant generation
have a relatively higher CPU time compared to their wall time
because these configurations spend some time generating invariants in parallel
to the \kinduction algorithm.
The results show, however,
that the time spent for the continuously-refined invariant generation
clearly pays off
as this configuration is not only the one with the most correct results,
but at the same time the fastest sound configuration with only~$170$\,h in total
($20$\,h~less than the second-fastest sound configuration).
The fact that the accumulated wall time ($9.4$\,h) it spent on correct results
is slightly higher than for most of the other sound configurations
is simply because it produced more correct results.
The accumulated CPU time ($17$\,h) spent on correct results
is higher than for most of the other configurations partly due to the same reason,
but also because of the multiple iterations
of the invariant-generation algorithm
as opposed to only one iteration
for the configurations using static invariant generation
or even zero iterations
for the configuration using no invariant generation
and the unsound configuration using the \esbmc heuristic.
Even though it produced much more correct results,
the configuration using continuous invariant generation
did not exceed the times of the
chosen configurations using static invariant generation ($>170$\,h).

These results show that the additional effort invested in
generating sound auxiliary invariants is well-spent,
as it even decreases the overall time due to the fewer timeouts.
As expected, the continuously-refined invariants
solve many tasks quicker than
the configurations using invariant generation with high static precisions.

\subsubsection{Final value of~$k$}
The bottom of Table~\ref{tab:induction-results} shows some statistics
about the final values of~$k$ for the correct safety proofs.
There is no difference between the maximum $k$~values
for the configuration using no auxiliary invariants,
the configuration~$(0,1,t)$ using low-precision invariants,
and the configuration~$(16,2,t)$ using medium-precision invariants.
The configuration using static invariant generation with high-precision
and the unsound configuration using the \esbmc heuristic
have higher maximum final values of~$k$,
with~$119$ for the high-precision configuration~$(16,2,f)$
and~$120$ for the unsound configuration.
The logs revealed that this unique deviation
of the high-precision static invariant-generation configuration
was caused by a situation where the static invariant generation
completed only shortly before the timeout ($k = 119$ instead of $k = 101$).
For the unsound configuration,
there was one case where due to the low overhead of the approach,
the iterative deepening of $k$ progressed quickly up until the value~$120$,
where the \kinduction proof then succeeded.
The configuration using continuously-refined invariants, on the other hand,
has a significantly lower maximum final $k$-value than the other configurations.
This is due to the following two reasons:
First, with continuously-refined invariants,
less time is wasted on generating unnecessarily strong invariants
than for static high-precision configurations,
and the proofs terminate before high values of $k$ are reached.
Second, the dynamicity of the approach
allows for generating stronger invariants than static low-precision configurations,
thus reducing the value of $k$ required for the proof to succeed.
%
%The average values of~$k$ are very similar for all configurations,
%with only two deviations.
%Both the unsound and the sound configuration using no auxiliary invariant generation
%have lower average values of $k$ than the others,
%which can be explained by the fact that due to the simplicity of the approaches
%and their lack of auxiliary invariants,
%these configurations are less likely to benefit from increasing $k$
%as compared to configurations that are able to rely on complex invariants.

\begin{table}
\centering
\caption{Results of \kinduction-based tools for~all~\nPrograms~verification tasks}
\label{tab:tool-results}
%\vspace{-1mm}
\begin{tabular}{l@{\quad}ffrf}
\hline
Tool & \multicolumn{1}{c}{\cbmc}      & \multicolumn{2}{c}{\esbmc}    & \multicolumn{1}{c}{\cpachecker} \\
Configuration       &     & \multicolumn{1}{c}{sequential} & \multicolumn{1}{c}{parallel} & \multicolumn{1}{c}{cont. refined} \\
\hline
	Score \phantom{\Large X} &       -971 &          1\,659 & 2\,027 & \textbf{3\,464}\\
	Correct results          &     1\,216 & \textbf{2\,214} & 2\,137 &          1\,981\\
	Wrong proofs             &        261 &             184 &    137 &      \textbf{1}\\
	Wrong alarms             & \textbf{4} &              28 &     24 &               7\\
	CPU time (h)             &        350 &             100 &    130 &             170\\
	Wall time (h)            &        350 &             100 &     76 &             100\\
\hline
\multicolumn{5}{l}{Times for correct results only: \phantom{\Large X}}\\
	CPU time (h)    & 1.9    & 34         & 25           &  17 \\
	Wall time (h)   & 1.9    & 34         & 14           & 9.4 \\
\hline
\multicolumn{5}{l}{$k$-Values for correct safe results only: \phantom{\Large X}} \\
	Max. final $k$  &  50    & 100        & 100          &  88 \\
	Avg. final $k$  & 1.1    & 8.4        & 7.4          & 2.4 \\
\hline
\end{tabular}
%\vspace{-2mm}
\end{table}

\subsection{Comparison with other tools}
\begin{figure}
\centering
\includegraphics[width=\linewidth]{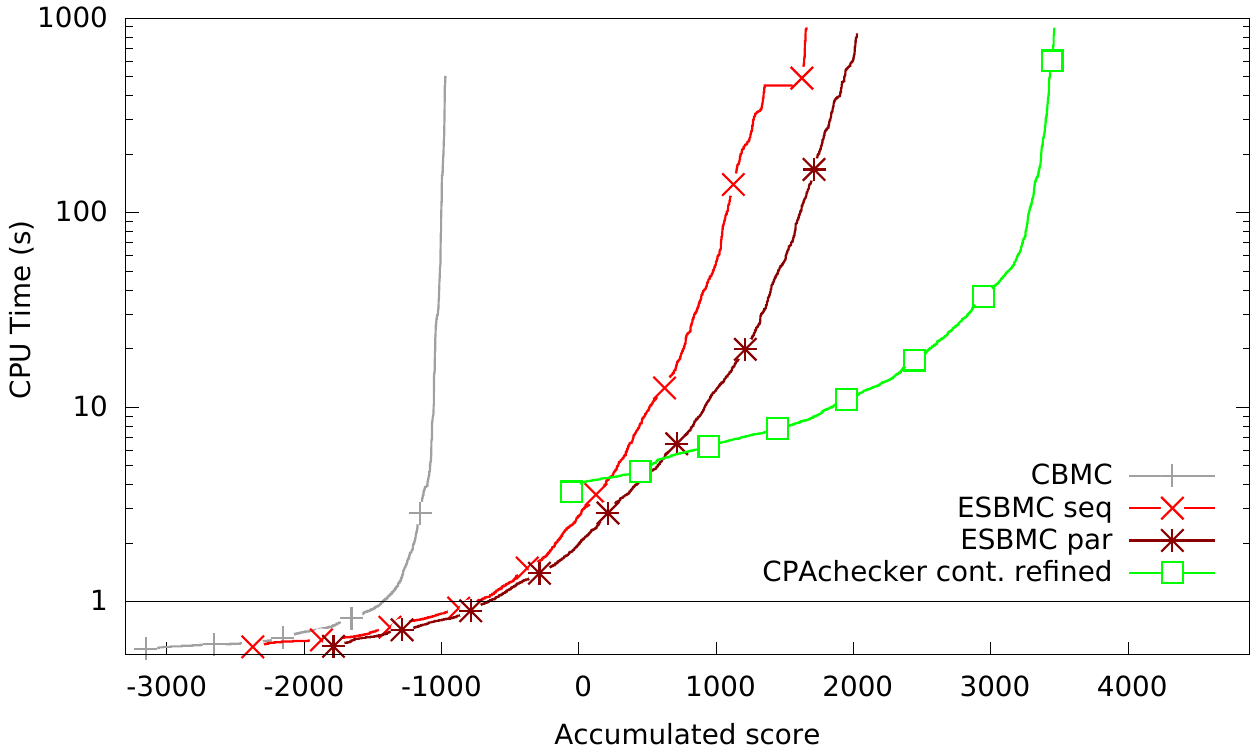}
%\vspace{-7mm}
\caption{Quantile functions of $k$-induction-based tools
for accumulated scores showing the CPU time for the successful results;
linear scale between 0\,s and 1\,s, logarithmic scale beyond}
\label{fig:experiments-quantile-cputime-scores-tools}
\vspace{-2mm}
\end{figure}

For comparison with other \kinduction-based tools,
we evaluated \esbmc and \cbmc,
two other successful software model checkers with support for \kinduction.
The \cpachecker configuration in this comparison
is the same as the one above
using continuously-refined invariants.
For \cbmc, we used the latest version~5.0
in combination with a wrapper script
for split-case \kinduction
provided by Michael Tautschnig.
For \esbmc we used the latest version~2.24.1
in combination with the wrapper script
of their submission to the 2013 Competition on Software Verification~\cite{ESBMC-COMP13}
(the script configures \esbmc to use \kinduction).
We also provide results for the experimental parallel \kinduction of \esbmc,
but note that our benchmark setup is not focused on parallelization
(using only two CPU cores and a CPU-time limit instead of a wall-time limit).
Table~\ref{tab:tool-results} summarizes the results;
Fig.~\ref{fig:experiments-quantile-cputime-scores-tools} shows the quantile functions
of the accumulated scores for each configuration.
The results for \cbmc are not competitive,
which may be attributed to the experimental nature of its \kinduction support.

\smallsec{Score}
Both configurations of \esbmc
produce a significant number of wrong results.
All tools do produce some wrong answers,
which are probably related to unsoundness and imprecision in the handling of some C~features.
\cpachecker with \kinduction and sound invariants has only 1~missed bug
(i.e, wrong claim of safety),
whereas \esbmc, in the sequential version, has 184~wrong safety proofs.
This large number of wrong results must be attributed to the unsound heuristic of \esbmc
for strengthening the induction hypothesis,
where it retains potentially incorrect information about loop-modified variables.
The large number of wrong proofs
reduces the confidence in the soundness of the correct proofs.
Consequently, the score achieved by \cpachecker
with continuously-refined invariants
is much higher than the score of \esbmc
(3\,464 instead of 2\,027 points).
This clear advantage is also visible in Fig.~\ref{fig:experiments-quantile-cputime-scores-tools}.

When comparing the results of \esbmc
to \cpachecker with a reimplemention of the unsound heuristic of \esbmc,
we see that \esbmc produces fewer wrong results.
The reason for this difference is that
the heuristic only works well
if relevant variables are identified on loop-exit conditions.
Due to \cpachecker's encoding of multiple loops in a program
into a single loop for \kinduction,
the number of loop-exit conditions is smaller than in the original program,
and the heuristic performs worse.
However, even with the implementation in \esbmc,
this unsound heuristic produces so many wrong results
that it is not suited for verifying program safety.

The parallel version of \esbmc performs somewhat better than its sequential version,
and misses fewer bugs.
This is due to the fact that the base case and the step case are performed in parallel,
and the loop bound~$k$ is incremented independently for each of them.
The base case is usually easier to solve for the SMT solver,
and thus the base-case checks proceed faster than the step-case checks
(reaching a higher value of~$k$ sooner).
Therefore, the parallel version manages to find some bugs
by reaching the relevant~$k$ in the base-case checks
earlier than in the step-case checks, which would produce a wrong safety proof at reaching~$k$.
However, the number of wrong proofs is still
much higher than with our approach, which is conceptually sound.
Thus, our score is more than 1\,400~points higher.

% \begin{figure}
% \centering
% \includegraphics[width=\linewidth]{experiments/scatter-cputime-esbmc.pdf}
% \caption{Scatter plot of the CPU time in seconds
% for all programs verified successfully by both \cpachecker and \esbmc (sequential) with $k$-induction}
% \label{fig:experiments-scatter-cputime-esbmc}
% \end{figure}

\smallsec{Performance}
Table~\ref{tab:tool-results} shows that,
if only the times for correct results are considered,
our approach is considerably faster than \esbmc
(\cbmc has so few correct results that the time for them is even less).
This indicates that due to our invariants,
we succeed more often with fewer loop unrollings
and thus in less time.
It also shows that the effort invested for generating the invariants
is well spent.
If considering the total time for the analysis of all results,
\cpachecker needs more time.
This is due to the fact that these measurements are dominated
by those programs for which the tool runs into a timeout,
and \cpachecker has more timeouts,
whereas \esbmc has more wrong results
(for which less time is spent).
A timeout is generally preferable to a wrong result, though.

\begin{figure}
\centering
\includegraphics[width=\linewidth]{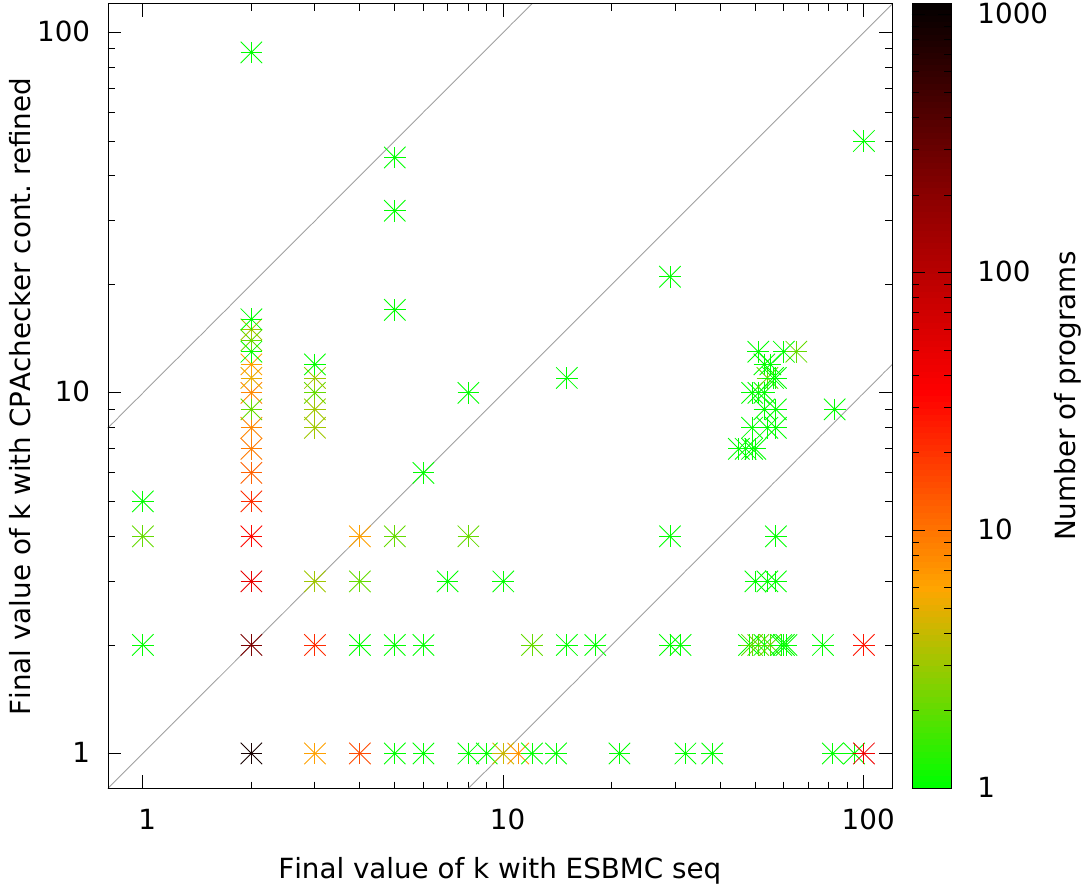}
%\vspace{-6mm}
\caption{Scatter plot of the final value of $k$
for all safe programs verified successfully by both \cpachecker (continuously-refined invariants) and \esbmc (sequential) with $k$-induction;
the color of each data point indicates the number of programs solved with this value of $k$}
\label{fig:experiments-scatter-k}
\vspace{-2mm}
\end{figure}

\smallsec{Final value of $k$}
The bottom of Table~\ref{tab:tool-results}
contains some statistics
on the final value of~$k$ that
was needed to verify a program.
Figure~\ref{fig:experiments-scatter-k} shows a scatter plot
comparing the values of the loop bound~$k$ for \cpachecker
with continuously-refined invariants
and \esbmc in its sequential version.
Both axes and the color range have a logarithmic scale.
Data points are shown only for those $1\,460$~verification tasks that can be proved safe
by both configurations.
A~point in the lower right half means
that \cpachecker needed a lower~$k$ (fewer loop unrollings)
than \esbmc for the same verification task.
The color of each data point gives an indication
of how many verification tasks are represented by the data point.
For example, the dark point at~$(2, 1)$
signifies that there are $845$~programs
that can be verified by \cpachecker
with a final value of~$k = 1$,
whereas \esbmc needs $k = 2$ for these programs.

The table shows that for safe programs, \cpachecker
only needs a loop bound
that is (on average) less than a third of the loop bound that \esbmc needs.
The bottom of the plot shows that there are many programs
(including the $845$~programs at $(2, 1)$)
that \cpachecker verifies with only one loop unrolling,
but for which \esbmc needs to unroll the loops more often.
To the right of the plot,
there is also a group of programs for which \esbmc
needs a~$k$ between $45$ and $65$ to verify the program,
and \cpachecker succeeds with significantly smaller~$k$.
There are only four programs for which \cpachecker needs a~$k$ larger than~$32$
(one program for $k=40$, $k=45$, $k=50$, and $k=88$ each).
For \esbmc, the largest number of loop unrollings is~$100$,
which is necessary for $71$~programs.
These advantages are due to the use of generated invariants,
which make the induction proofs easier
and likely to succeed with a smaller number of~$k$.
There is also a group of programs where \esbmc
succeeds with $2$~loop unrollings
but \cpachecker needs up to~$16$.
However, the number of such programs is relatively small
(note that the data points with a green-to-orange color
only represent $1$~to $9$~programs)
and there is only a single program where \cpachecker
unrolls the loops more than $10$~times more than \esbmc
(while there are many with the reverse being true).
The reason why \esbmc needs fewer loop unrollings
for some programs is its (unsound) heuristic
of keeping information about some program variables
from the initial program state in the inductive-step case.

% \begin{figure}
% \centering
% \includegraphics[width=\linewidth]{experiments/scatter-cputime-predicate.pdf}
% \caption{Scatter plot of the CPU time in seconds
% for all programs verified successfully by \cpachecker both with $k$-induction and predicate abstraction}
% \label{fig:experiments-scatter-cputime-predicate}
% \end{figure}

\begin{figure}
\centering
\includegraphics[width=\linewidth]{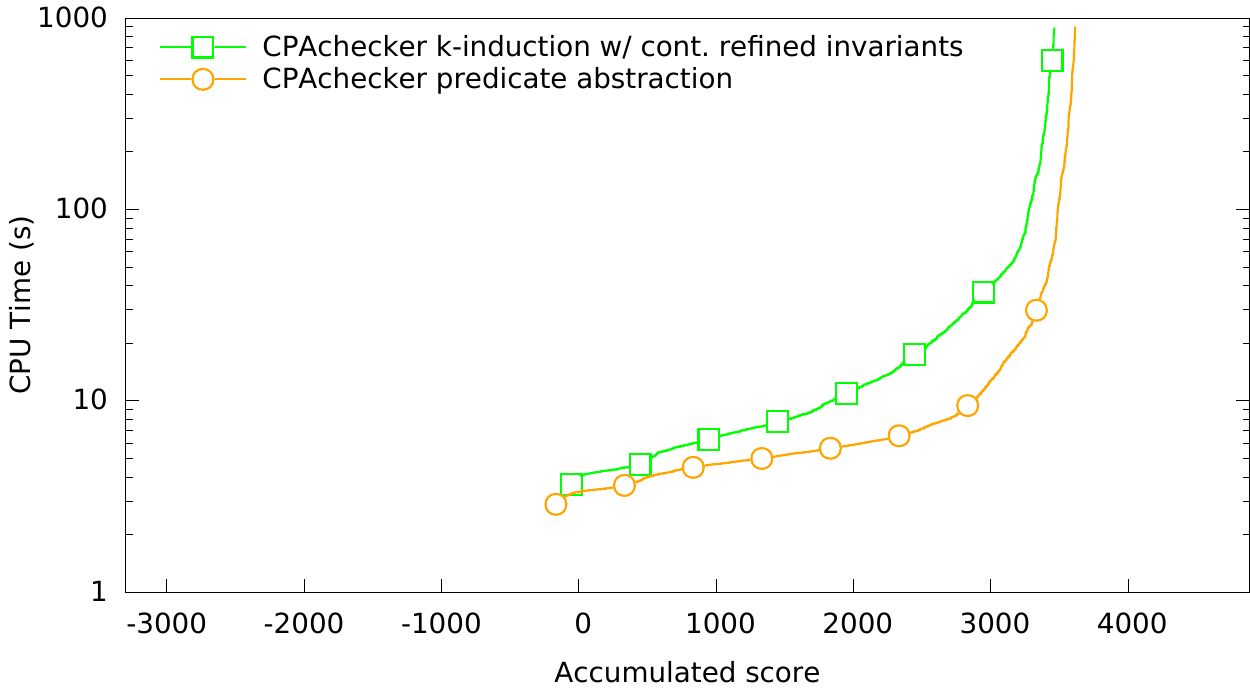}
%\vspace{-5mm}
\caption{Quantile functions
for accumulated scores showing the CPU time for the successful results}
\label{fig:experiments-quantile-cputime-scores-other}
\vspace{-2mm}
\end{figure}

\subsection{Comparison with other approaches}
We also compare with the predicate-abstraction implementation of \cpachecker~\cite{ABE},
which uses the same framework (parser, formula encoding, etc.) and SMT solver
as our implementation of \kinduction.
The score-based quantile functions
for our \kinduction approach
and the existing predicate abstraction
in Fig.~\ref{fig:experiments-quantile-cputime-scores-other}
show that the latter is somewhat faster and achieves a higher score.
It is surprising that even the well-tuned \cpachecker implementation
of the mature predicate-abstraction approach
only slightly outperforms our novel \kinduction implementation.
The difference in performance and score between these two configurations
is much smaller than the improvement of our approach
compared to existing \kinduction-based approaches
(cf. Fig.~\ref{fig:experiments-quantile-cputime-scores-tools}).
This is a promising result,
considering that there is room for improvement
in our approach.
Especially the invariant generation could be further enhanced,
e.g., by tailoring the invariant generation to the special needs
of the \kinduction proof,
and a more targeted invariant-refinement procedure.

\subsection{Acknowledgments}
We would like to thank M.~Tautschnig
and L.~Cordeiro
for explaining the optimal available configuration for \kinduction,
for the verifiers \cbmc and \esbmc, respectively.

\section{Conclusion}
We have presented the novel idea
of combining \kinduction with continuously-refined invariants,
and contribute a publicly available implementation of our idea
within the software-verification framework \cpachecker.
Our extensive experiments show
that our approach outperforms all existing implementations of \kinduction
for software verification, and that it is competitive
compared to other, more mature techniques for software verification.
We showed that a sound, effective, and efficient \kinduction approach
to general purpose software verification is possible,
and that the additional resources
required to achieve these combined benefits
are negligible if invested judiciously.
At the same time,
there is still room for improvement of our technique.
In the future, we plan to integrate successful features
of other approaches to \kinduction{}
such as the parallel algorithm of \esbmc.
The experiments with \esbmc show that we can avoid more timeouts on unsafe programs
by running
the iteratively-deepening BMC
decoupled from the slower inductive-step case.
We are also interested
in adding an information flow
between the two cooperating algorithms
in the reverse direction.
If the \kinduction procedure could tell
the invariant generation which facts
it misses to prove safety,
this could lead to a more efficient and effective approach
that generates invariants that are specifically tailored
to the needs of the \kinduction proof.
Already now, \cpachecker is parsimonious in terms of unrollings,
compared to other tools.
The low $k$-values required to prove many programs
show that even our current invariant generation is powerful enough
to produce invariants that are strong enough
to help cut down the necessary number of loop unrollings.
\kinduction-guided precision refinement might direct the invariant generation towards
providing weaker but still useful invariants
for \kinduction more efficiently.

%\def\IEEEbibitemsep{1pt plus .5pt}

%\clearpage % ICSE allows up to 2 separate pages for bibliography

\balance
\bibliography{sw,dbeyer,tah,bib}

\end{document}